\documentstyle[cjaa209,twoside,epsf]{article}
\input{psfig.sty}                               

\begin{document}

\title{Keys to Cosmology -- Clusters of Galaxies }

\author{Sabine Schindler$^{1}$\mailto{}}

\inst{
$^1$ Institut f\"ur Astrophysik, Universit\"at Innsbruck,
Technikerstr. 25, A-6020 Innsbruck, Austria\\
}

\email{Sabine.Schindler@uibk.ac.at}

\markboth{Sabine Schindler}
{ Keys to Cosmology -- Clusters of Galaxies }

\pagestyle{myheadings}

\date{Received~~2003~~~~~~~~~~~~~~~ ; accepted~~2003~~~~~~~~~~~~~~ }

\baselineskip=12pt


\begin{abstract}
\baselineskip=12pt
We review several aspects of clusters of galaxies and their
application to cosmology. 
We present first results of numerical simulations of the
dynamics of the intra-cluster gas and of different interaction
processes between cluster galaxies and the intra-cluster gas. In
particular metallicity maps are very useful to determine the
importance of the different interaction processes.
Also mass determination methods and possible sources for uncertainties in the
measurements are shown. 


\end{abstract}

%
%
\section{Introduction}           

Clusters of galaxies are very versatile tools for various types of
analyses. They can be used to determine cosmological parameters as
well as physical processes on large scales and extreme
environments. Out of this large field we select here mass and dark
matter measurements, determination of the dynamical state and the
investigation of the interaction between the cluster galaxies and the
intra-cluster gas.

\section{Mass Determination and Dark Matter}

As clusters of galaxies are large structures which can be
regarded as being 
representative for the universe as a whole
their ratio of baryonic matter to dark matter is a crucial
number. Dark matter makes up  about 80\% of the total cluster mass,
with the total cluster mass being determined by different methods. 

One
of the methods uses X-ray observations. The X-ray emitting gas
traces the total cluster potential and hence traces the total cluster
mass. With this mass determination one finds the following mass
fractions: mass in galaxies 3-5\%, mass in the intra-cluster gas 15-20\%.

For the X-ray mass determination two assumptions are required: hydrostatic
equilibrium and spherical symmetry. Tests with a sample of X-ray clusters
together with analytical models showed that even if clusters are not
spherically symmetric, but elongated the determination of total masses
and gas mass fractions is quite reliable (Piffaretti et al. 2003). Only
if projected clusters masses are calculated for the comparisons with
masses from gravitational lensing differences are visible (see Fig.~1).
Therefore
part of the observed discrepancies between X-ray masses and lensing
masses can come from these projection effects.

\begin{figure}
   \vspace{2mm}
   \begin{center}
   \hspace{3mm}\psfig{figure=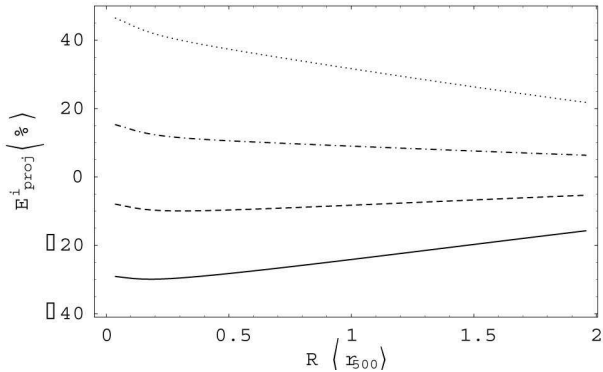,width=80mm,angle=0.0,clip=}
   \parbox{180mm}{{\vspace{2mm} }}
   \caption{The relative errors $E^i_{proj}(R)$ for the projected mass
   estimates in a cluster plotted versus the radius 
   for 4 triaxial models: compressed along
   the line of sight (dotted line), prolate (dash-dotted line), oblate
   (dashed line) and elongated along the line of sight 
   (solid line). Negative relative
   errors imply underestimates of the mass, if spherical symmetry is
   assumed. This is the case if the cluster is
   elongated. Overestimates are found for compressed
   clusters. Therefore an elongation along the line of sight can
   contribute to resolve the discrepancy between X-ray and lensing
   mass (from Piffaretti et al. 2003).}
   \label{Fig:lightcurve-ADAri}
   \end{center}
\end{figure}

The second assumption, hydrostatic equilibrium, was tested by
hydrodynamic simulations and found to be a valid assumption for
relaxed i.e. non-merger clusters (Evrard et al. 1996; Schindler 1996).
Clusters in the process of a major merger are obviously not in
hydrostatic equilibrium and therefore the mass determination can
easily be wrong by a factor of two.

Another possible source for the discrepancy between X-ray and lensing
masses found in some clusters can be non-thermal pressure. While in the
centre of clusters the negligence of non-thermal pressure from cosmic rays and magnetic
fields can lead to a significant mass underestimate (Colafrancesco et
al. 2003), we found that in
the outer part, where the mass is usually determined, the non-thermal
pressure from magnetic fields is negligible (Dolag \& Schindler
2000). Only if the clusters are in the process of merging, their magnetic
field can temporarily be so strongly enhanced that a mass underestimate down to
50\%  can be found.

A second way to determine the mass is gravitational lensing. Light
from background galaxies is deflected by the huge mass of a
cluster. Therefore the images of these background galaxies are
distorted and show up as arcs (= strong lensing). The arcs are very
thin and typically very faint structures. Under
non-ideal observing conditions (e.g. bad seeing) 
they are easily dispersed and disappear
into the background. Even under ideal conditions they are not easy to
detect, because they are often just above the background level. To
remove the noise and make faint structures better visible usually
smoothing is applied. Unfortunately in the case of such thin
structures as arcs the smoothing procedure often leads to a dispersion of
the few photons so that the arcs are not detectable.
To prevent this dispersing we have developed an algorithm that
automatically smooths only along the arcs and not perpendicular to
them, by so-called ``anisotropic diffusion'' 
(see Fig.~2, Lenzen et al. 2003). The subsequently applied
source finding procedure extracts all the information from the sources
necessary to distinguish arcs from other sources. This new algorithm
is much more efficient in finding gravitational arcs than existing
source detection algorithms  because it is
optimised just for this purpose. 

\begin{figure}
   \vspace{2mm}
   \begin{center}
   \hspace{3mm}\psfig{figure=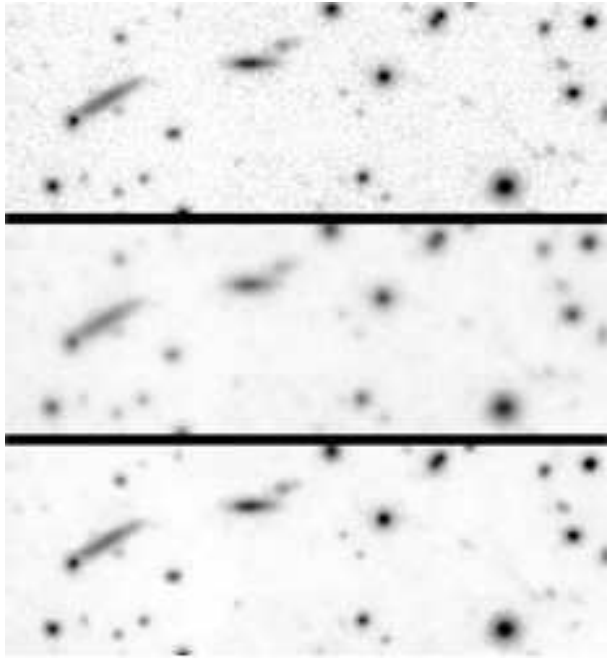,width=80mm,angle=0.0,clip=}
   \parbox{180mm}{{\vspace{2mm} }}
   \caption{Comparison of different smoothing methods. Top: original
   image. Middle: Image smoothed with a Gaussian filter. Bottom: Image
   smoothed with the newly developed algorithm of anisotropic
   diffusion. In the Gaussian filtered image (middle) the edges are
   not well preserved, i.e. the arcs get dispersed, while anisotropic
   dispersion (bottom) maintains the edges and reduces the noise at
   the same time
   (from Lenzen et al. 2003).}
   \label{Fig:lenzen}
   \end{center}
\end{figure}

\begin{figure}
   \vspace{2mm}
   \begin{center}
\psfig{figure=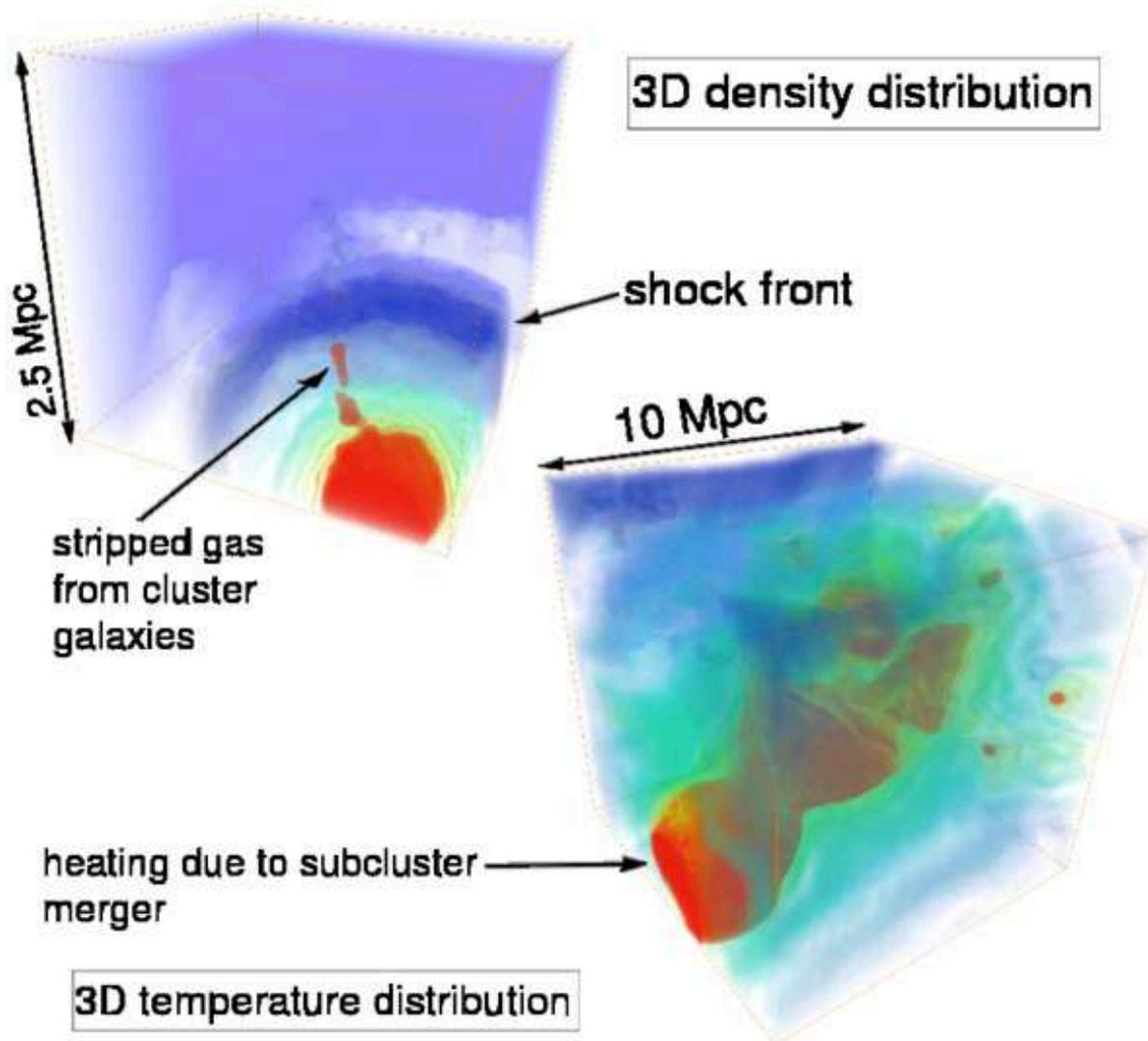,width=160mm,angle=0.0,clip=}
   \parbox{180mm}{{\vspace{2mm} }}
   \caption{Density and temperature distribution of the intra-cluster
   gas. Mergers of subclusters lead to shock fronts (left) and to
   heated regions in particular between the subclusters just before
   the collision (right). Note the different scale of the two images
   (Visualisation by W. Kapferer).}
   \label{Fig:lenzen}
   \end{center}
\end{figure}

The matter density of the universe $\Omega_m$ can be determined 
with mass determinations of galaxy clusters in
two ways. One possibility is to measure total mass and baryonic mass
and assume that the ratio between the two is representative for the
universe as a whole. This methods yields values around $\Omega_m
\approx 0.3$ (Ettori et al. 2003,
Castillo-Morales \& Schindler 2003). Another way is to determine the
mass function which yields similar results (Rosati et al. 2002,
Reiprich \& B\"ohriger 2002). In the latter method
one must be
careful with selection effects and source confusion (see 
Gil-Merino \& Schindler 2003).

\begin{figure}
   \vspace{2mm}
   \begin{center}
\psfig{figure=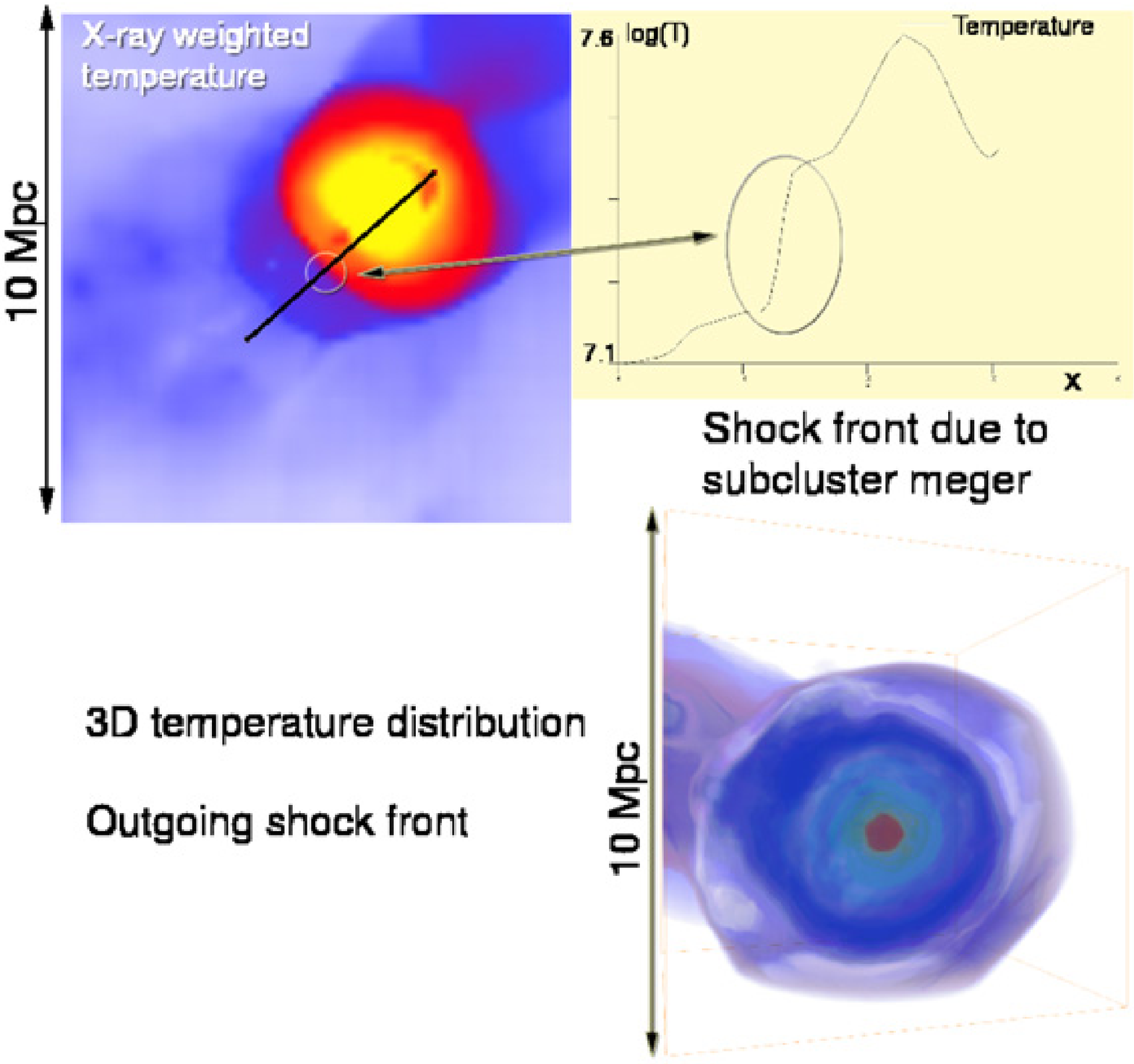,width=160mm,angle=0.0,clip=}
   \parbox{180mm}{{\vspace{2mm} }}
   \caption{Shocks emerging from subcluster mergers are seen as steep
   gradients in the temperature distribution. Top: Temperature map as
   it would appear in an X-ray observation of a cluster shortly after
   a collision.  For
   a quantitative view a trace through the shock is shown on the
   right. Bottom:  3D view of an outgoing shock after the collision of
   subclusters
   (Visualisation by W. Kapferer).}
   \label{Fig:lenzen}
   \end{center}
\end{figure}

\section{Dynamical State}

The dynamical state of clusters depends sensitively on the matter
density and slightly also on the dark energy. Therefore the
determination of the dynamical states of many clusters is another
independent way
to constrain cosmological parameters.

As from observations one can obtain only snapshots of the cluster
evolution, it is ideal to use numerical simulations to learn how
different dynamical states appear in X-ray images or temperature
maps. Therefore we perform hydrodynamic simulations of the
intra-cluster medium in a collaboration between University of
Innsbruck
(Domainko, Kapferer, Kimeswenger, Mair, Schindler, van Kampen),
University of Edinburgh (Mangete, Ruffert) and Max-Planck-Institut
f\"ur Gravitationsphysik (Benger). We use a grid code (Piecewise
Parabolic Method, Colella \& Woodward 1984) with multiply nested 
grids (Ruffert 1992).

Preliminary results of the simulations are shown in Figs.~3 and 4. 
Subcluster mergers
lead to numerous shocks, which are moving outwards. 
The shocks appear in the temperature as well as in the
density distribution as steep gradients. It is possible that
particles, which are responsible for the radio haloes in several clusters,
are accelerated to relativistic energies in these shocks (Giovannini
\& Feretti 2002).

Even before the collision of
subclusters the gas is heated. The gas between the two subclusters is
compresses and therefore it is 
heated. Hence a hot region 
between the subclusters is visible in the temperature map of the
intra-cluster gas. 
For more information, images and films see 
http://astro.uibk.ac.at/astroneu/hydroskiteam/index.htm.

Such a feature of a heated region is e.g. observed in the cluster CL0939+4713
(De Filippis et al. 2003). The X-ray image shows (apart from several
point-like sources) two extended subclusters. The temperature map clearly
shows a hot region between the two subclusters as it is expected for the
early stage of an (almost) central collision (see Fig.~5).

\begin{figure}
   \vspace{2mm}
   \begin{center}\hspace{2cm}
   \psfig{figure=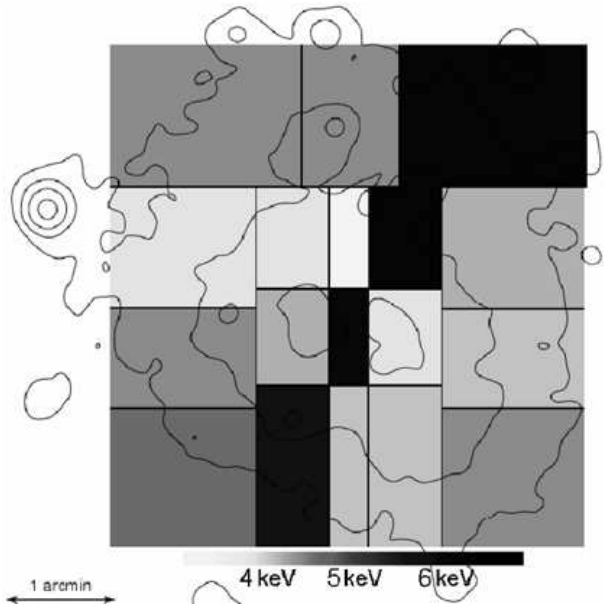,width=80mm,angle=0.0,clip=}
   \parbox{180mm}{{\vspace{2mm} }}
   \caption{XMM observation of the cluster CL0939+4713. The contours
   show the X-ray surface brightness distribution. Superposed on it is
   the temperature map in grey scales with the hotter gas being
   darker. Between the two subclusters is a hot region which implies
   that they are approaching each other (from De Filippis et al. 2003).}
   \label{Fig:cl0939}
   \end{center}
\end{figure}

\section{Interaction between Galaxies and the Intra-Cluster Gas}

Until recently galaxies and the intra-cluster gas have been treated
mostly independently. 
But in the last years a lot of examples of
interaction between these two components have been found. The evidence
does not only come observations of
galaxies in several wavelengths, but also from observations of the
intra-cluster gas, in particular of the metal abundances. 
As the metal content of the intra-cluster gas is
on average roughly one third of the solar value, i.e. the total amount
of metals in the intra-cluster gas is about the same as in all
galaxies together, a lot of gas must have been
transported from the galaxies into the intra-cluster medium. Several
transport processes have been suggested: ram pressure stripping (Gunn
\& Gott 1972), galactic winds (De Young 1978), galaxy-galaxy
interaction, jets from AGN and others.
A closer look at the different metal enrichment processes is therefore very
important for the understanding of cluster formation and galaxy
evolution. So far numerical simulations including only some of the
processes gave quite discordant results
on the efficiency of the different processes (e.g. Cen \& Ostriker
1999; Aguirre et al. 2001; Metzler \&  Evrard 1994, 1997). To improve
on this we have
started a comprehensive project. We use the hydrodynamic simulations mentioned
above and include all possible
enrichment processes. We calculate simulated metallicity maps for direct
comparison with observations.

Finally, the first observed metallicity maps are 
available because
the X-ray satellites XMM and CHANDRA have sufficiently high
sensitivity and provide the possibility for spatially resolved
spectroscopy. Therefore the evolution of metals in the intra-cluster
gas as well as the spatial distribution can be measured now and be
compared with simulations. Recently, in a number of clusters
variations in the overall metallicity distribution have been found -- not only in the
cluster centre 
(Perseus cluster: Schmidt et al. 2002, 
A2199: Johnstone et al. 2002,
A3558 and 3C129: Furuzawa et al. 2002, 
A1060: Yamasaki et al. 2002,
A3671: Hudaverdi et al. 2002,
2A0335+096: Tanaka et al. 2002, 
AWM7: Furusho 2002,
Cl0939+4713: De Filippis et al. 2003).

\begin{figure}
   \vspace{2mm}
   \begin{center}
\psfig{figure=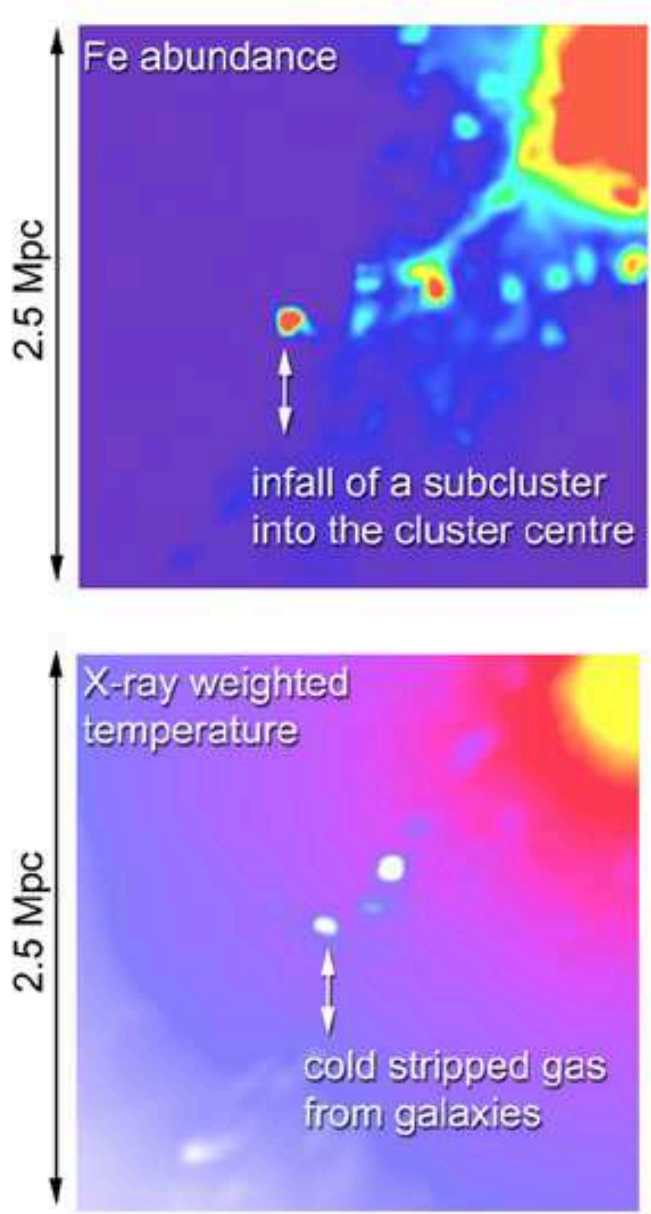,width=150mm,angle=0.0,clip=}
   \parbox{180mm}{{\vspace{2mm} }}
   \caption{As galaxies and subclusters fall onto the centre of a
   cluster they feel more and more pressure of the intra-cluster
   gas. At some point this can lead to ram-pressure stripping. The
   galaxies lose their cool, metal-enriched gas to the intra-cluster
   gas. Here we show the distribution of the iron abundance and the
   temperature in a simulated cluster. 
   The centre of the cluster is in the upper right corner. 
   Top: regions of high iron abundance are visible, where
   subclusters and galaxies have been stripped. Bottom: the cool gas
   stripped off galaxies is visible for a while in the temperature
   map. It is surrounded by the
   hotter intra-cluster medium
   (Visualisation by W. Kapferer).}
   \label{Fig:lenzen}
   \end{center}
\end{figure}

First results of the simulations
are shown in Figs.~6 and 7. The simulations shown here include so far only 
ram-pressure stripping. While many other simulations have concentrated
on the effect of the stripping on the galaxies
(Abadi et al. 1999;
Mori \& Burkert 2000;
Quilis et al. 2000;
Vollmer et al. 2001;
Schulz \& Struck 2001;
Toniazzo \& Schindler 2001;
Otmianowska-Mazur \& Vollmer 2003), we concentrate here on
the effects of the stripping process on the intra-cluster gas.

The gas that is stripped off the galaxies is
visible not only in the metallicity maps of the intra-cluster gas, but also 
in the temperature maps, because this gas is cooler than the
surrounding intra-cluster medium. It takes a while for the gas from
the galaxies to mix
with the intra-cluster gas. Sometimes one can see even shock waves
passing over these cool regions and the cool gas is not heated up
immediately. Also the metallicity
distribution is not expected to be homogeneous shortly after the
stripping, because the gas is not mixed immediately.
This expectation
is indeed supported be the metallicity variations observed
in the metallicity maps mentioned above. In the simulations we 
see the highest stripping rate
and hence the highest metallicities in regions, where subclusters fall
onto the main cluster.

For more information on the simulation method, first results in form
of images and films see 
http://astro.uibk.ac.at/astroneu/hydroskiteam/index.htm.

\begin{figure}
   \vspace{2mm}
   \begin{center}
\psfig{figure=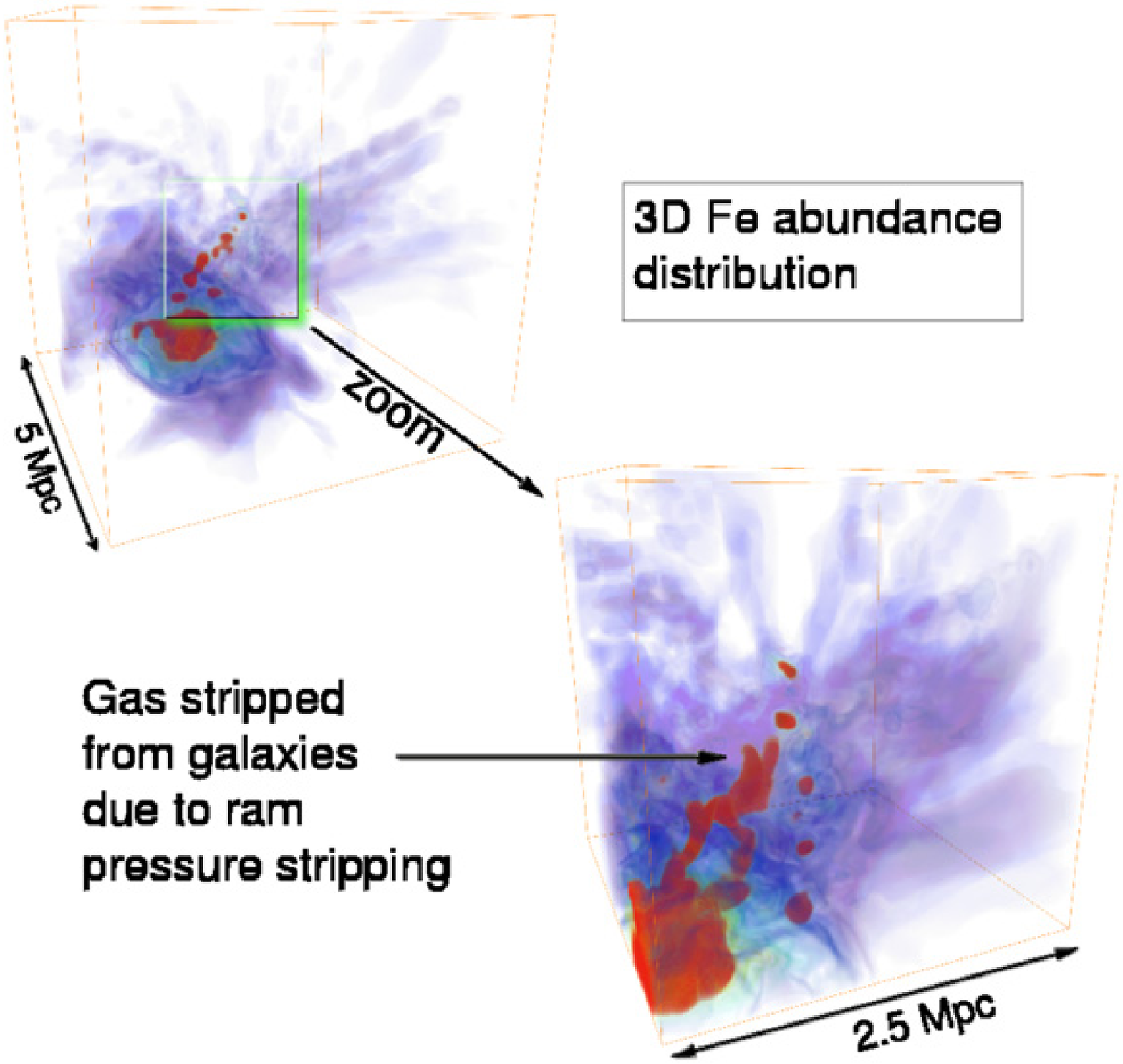,width=160mm,angle=0.0,clip=}
   \parbox{180mm}{{\vspace{2mm} }}
   \caption{3D distribution of the iron abundance in the ICM 
   after the ISM has been
   stripped off the galaxies. Two different scales (5 Mpc and 2.5 Mpc)
   are shown  
   (Visualisation by W. Kapferer).}
   \label{Fig:lenzen}
   \end{center}
\end{figure}


\begin{acknowledgements}
This work is supported partly by the Austrian Science Foundation FWF,
Project P15868, and by the EU Exchange Programme ``Training and
Research on Advanced 
Computing Systems'' (TRACS) between Innsbruck and Edinburgh. 
\end{acknowledgements}

\end{document}